\documentclass[twocolumn,showpacs,preprintnumbers]{revtex4}

\usepackage{amssymb}
\usepackage{graphicx}
\usepackage{dcolumn}
\usepackage{bm}
\usepackage{dsfont}
\usepackage{amsmath}
\usepackage{amsthm}
\usepackage{amscd}

\begin{document}

\title{Rotation of the swing plane of Foucault's pendulum and Thomas spin \\precession: Two faces of one coin.
}

\author{M. I. Krivoruchenko\footnote{
On leave of absence from Institute for Theoretical and Experimental Physics, B. Cheremushkinskaya 25, 117218 Moscow, Russia
}
}
\affiliation{
Institut f\"{u}r Theoretische Physik$\mathrm{,}$ T\"{u}bingen Universit\"{a}t$\mathrm{,}$ Auf der Morgenstelle 14\\ 
D-72076 T\"{u}bingen$\mathrm{,}$ Germany
}

\begin{abstract}
Using elementary geometric tools, we apply essentially the same methods to derive expressions 
for the rotation angle of the swing plane of Foucault's pendulum 
and the rotation angle of the spin of a relativistic particle moving in a circular orbit 
(Thomas precession effect).
\end{abstract}

\maketitle


\section{Introduction}

Jean Bernard L\'eon Foucault conducted his first pendulum experiment in Paris in January 1851, aiming to prove the rotation
of the Earth by demonstrating the rotation of the plane of swings. Originally, the suspension length of the pendulum was 2 m. 
The next experiment was set up with the suspension length 11 m at the Observatory of Paris. 
Louis-Napol\'eon Bonaparte,
the first titular President of the French Republic and nephew of the famed Napoleon I of France,
was informed of Foucault's work and proposed him to conduct an experiment at the Panth\'eon. The
experiment took place on 31 March 1851, with the pendulum bob weight 28 kg suspended under the Panth\'eon dome by a
steel wire 67 m long.

Aristarchus of Samos who proposed the first consistent heliocentric model around 270 B. C. 
explained the observed rotation of stars by the axial rotation of the Earth. Similar ideas were 
discussed earlier by Philolaus, philosopher of the Pythagorean school, 
in the V-th century B. C. and by Greek philosopher Herakleides 
in the IV-th century B. C.
If idea on the axial rotation of the Earth is correct, one can expect
that because of inertia the plane of swings of the Foucault's pendulum does
revolve delaying the Earth rotation.
It, however, the Earth is fixed and motionless, as was embraced by both
Aristotle, Ptolemy, and most Greek philosophers, the
swing plane of Foucault's pendulum cannot rotate. 

From the technical point of view, the experiment with the Foucault's pendulum
was accessible to all ancient and more recent civilizations, including Greeks, 
nevertheless, the experiment has been done in the New Era only. During one and half 
thousands of years it was assumed that the problem of rotation of the stars does 
not require additional attention because of the incontestable authority of Aristotle
and success of the Ptolemy heocentric model that described and describes 
until now the motion of planes with high precision. The interest to the problem 
and discussions recommenced again in the XVI century after the works of Nicolaus Copernicus
and completed essentially after the works of Johannes Kepler at the beginning 
of the XVII century.

From the viewpoint of an observer at the Earth, remote stars complete 
one rotation in the clockwise direction during 1 sidereal day = 23 hours, 
56 minutes, 4.091 seconds. 

The observed rotation rate, $\dot {\varphi}_{E} \approx - 11^{\circ}$
per hour, of Foucault's pendulum is not equal to $- 360^{\circ}/24 = -
15^{\circ}$ per hour, nor is it zero (the minus sign indicates that the
swing plane rotates clockwise).

If an observer in the coordinate system of the remote stars transports the Foucault's pendulum to the North Pole along 
the Earth meridians keeping the angle between the pendulum's plane and the meridians fixed, she will
detect a uniform precession of the pendulum's plane relative to the remote stars. The rotation angles 
$\varphi_{S}$ and $\varphi_{E}$, relative to the remote stars and the Earth meridians, respectively, 
are related by $\varphi_{S} = 2\pi + \varphi_{E}$ for one sidereal day. In the adiabatic approximation 
and for small swing angles, the rotation angle $\varphi_{S}$ is given by
\begin{equation}
\varphi_{S} = 2\pi (1 - \cos \vartheta),
\label{ODINFOUC}
\end{equation}
where $\vartheta$ is the polar angle. The adiabaticity means that the period of the Earth's rotation is much greater 
than the period of the swings.

The equation of rotation of the plane of swings of Foucault's pendulum, as illustration of the 
laws of classical mechanics (see, e.g., \cite{arnold} and also Section III), is an element of 
program of university courses of physical faculties. 

Thus, the Foucault's pendulum, placed at the North Pole, 
during the day turns at an angle $\varphi_E = - 360 ^{\circ}$. 
At the Equator the pendulum does not rotate. The Panth\'eon
in Paris is located at the parallel $\vartheta = 41.15 ^{\circ}$ (in geography 
one speaks about the latitude $ \alpha = \pi /2 - \vartheta $). 
From the equations we obtain $ \dot{\varphi}_E = - 11.3^{\circ}$ per hour, which is consistent 
with the observations on the Foucault's pendulum and precludes accompanying "rotation of heavens"
with high accuracy.

The rotation of the plane of the pendulum provided the first evidence for the 
rotation of the Earth by terrestrial means.

In the process of movement of the pendulum along the surface of the Earth the plane of swings, 
as a consequence of the laws of classical mechanics, remains parallel to itself 
\cite{SOME,HART,WILC,JENS,BERR}. This remarkable fact allows to investigate the evolution of 
the Foucault's pendulum using geometric methods.

The bob's velocity in the equilibrium point is tangent to
the surface of the Earth, and therefore belongs to the tangent space of the surface.
The plane of swings of the pendulum can be described by a vector orthogonal to it. 
Such a vector lies in a plane tangent to the Earth's surface also. 
In the absence of external forces and/or torques tangent vectors under the displacement
experience \textit {parallel transport}. For example, vectors tangent to Minkowski spacetime, 
$\mathbb{M}=\mathbb{R}^{1,3}$, like four-velocity, remain fixed under displacement, 
whereas tangent vectors of curved spaces like the surface of a sphere, $\mathbb{S}^2$, 
or the space of physical relativistic velocities, rotate under displacement in order 
to remain tangent to the corresponding surface. This kind of evolution is associated 
with \textit{inertial motion}. In the first case the displacements occur in Minkowski 
spacetime, in the second and third cases the displacements occur on the surface of a sphere 
and in the space of physical relativistic velocities, respectively.

The geometric basis of the relativistic effect known today as Thomas precession has been discovered in 1913 year by French 
mathematician \'Emile Borel \cite{BORE13}. He pointed out noncommutativity of non-collinear 
Lorentz transformations, described an analogy between transformations of vectors on a spherical surface 
and in the physical relativistic velocity space, and provided the lowest order estimate 
of kinematic precession of axes of a rigid body on a circular orbit.

The same year two young mathematicians Ludwig F\"oppl and Percy John Daniell derived exact 
expression for the rotation angle due to Thomas precession, corresponding to one period of 
a uniform circular motion \cite{FOPP13}
\begin{equation}
\phi_{S} = 2\pi (1 - \cosh \theta),
\label{ODINTHOM}
\end{equation}
where $\theta$ is rapidity, $\cosh \theta = \gamma = (1 - v^2/c^2)^{-1/2}$ is Lorentz factor, $v$ is velocity 
of the body, and $c$ is speed of light. The work of F\"oppl and Daniell 
has been recommended for publication by David Hilbert.

The relativistic precession effect has been known at about the same time to Ludwik Silberstein \cite{SILB14}.

In the early 1920s, Enrico Fermi \cite{FERM}, and later Arthur Walker \cite{WALK}, established a transport rule for vectors to construct preferred reference frames in the general theory of
relativity. In the Fermi-Walker transport, vectors tangent to the physical relativistic velocity space behave 
like the axes of a rigid body as described by Borel et al. \cite{BORE13,FOPP13,SILB14}.

The problem of the relativistic precession has attracted attention of physicists,
since Llewellyn Thomas \cite{THOM26} uncovered its fundamental significance for theory of atomic spectra. 
The effect was found to reduce the spin-orbit splitting of the atomic energy 
levels by a factor 1/2 known presently as the Thomas half. In the course of his work,
Thomas has only been aware of de Sitter's paper on the relativistic precession of the Moon, published in a book by 
Arthur Eddington \cite{EDDI24}.  

Group-theory aspects of the spin rotation under the Lorentz transformations were introduced to physicists 
by Eugene Paul Wigner \cite{WIGN39}. The term "Wigner rotation" is used as a synonym of the rotation of a spinning particle 
under the coordinate transformations and specifically as a synonym of "Thomas precession", too.

The history of early study of the relativistic precession effect is described in Ref. \cite{WALT99}.
As claimed by Robert Merton, multiple independent discoveries represent the common pattern in science \cite{MERT}.

Recently, the geometric nature of the spin precession effect of a relativistic particle has again attracted attention.
It was shown in \cite{ARAV97} that the rotation angle $\phi_S$ is determined by an integral over the surface limited by 
the closed trajectory of the particle in the physical relativistic velocity space. This property characterizes parallel transport of vectors in a
Riemannian space (see, e.g., \cite{LANDLI}). Parallel transport in the relativistic velocity space and Thomas precession 
were discussed in \cite{RHOD} in detail.

Rotation angles $\varphi_S$ and $\phi_S$ correspond to a geometrical phase that occurs in many areas of physics [4, 5, 18].

The analogy between rotations and Lorentz transformations is of the high heuristic value. In particular, 
we recall that the relativistic velocity addition theorem can be obtained
as the composition law for arcs of the great circles on a sphere
of imaginary radius in a four-dimensional Euclidean space
with one imaginary coordinate (time). By introducing the
imaginary coordinate (time), it is possible to transform a
hyperboloid of physical relativistic velocities into a sphere of
imaginary radius in a four-dimensional Euclidean space.
Both for Foucault's pendulum and for Thomas precession, 
the surface along which a displacement occurs can be considered a sphere. 
This suggests that the rotation effects of Foucault's swinging pendulum and
Thomas precession, obviously, are geometrically identical; this does not necessarily contradict to the different physical
natures of these systems.

The aim of this methodological note is to show that expressions for the rotation angles $\varphi_S$ and $\phi_S$  can be
obtained by the same method using elementary geometrical tools for parallel transport of vectors over corresponding
surfaces. In the first case, this is the Earth's surface, i.e., a spherical surface in the Euclidean space $\mathbb{R}^3$. In the second
case, this is the physical relativistic velocity space, i.e., the
hyperboloid $u^2=  1$ in the tangent space $T_x \mathbb{M}$ of Minkowski spacetime.

The visual tangent-cone geometric method used to derive the main equations in Sections III and IV is often used to
illustrate the effect of curvature on the parallel transport of vectors along a spherical surface (see, e.g., Appendix 1 in \cite{arnold}).
This method was used by Somerville \cite{SOME} and Hart, Miller, and R. L. Mills \cite{HART,COMM} to describe the evolution
of Foucault's pendulum. In Section II, we recall the main principles of parallel transport. In Section III, based on the
consideration of Foucault's pendulum evolution from the dynamic standpoint, we show that the evolution is reduced to
parallel transport of the swing plane of the pendulum over a spherical surface, and obtain Eq. (1). In Section IV, the
tangent-cone method is generalized to the case of Thomas precession and is used to obtain expression (2).

\section{Parallel transport}

\subsection{Euclidean space}

The concept of parallel transport of vectors originates in Euclidean geometry. Two vectors 
are said to be parallel if two straight lines that pass through the endpoints of these 
vectors are parallel in the sense of Euclid's fifth Postulate and their orientations coincide.
Any continuous transformation which preserves length and keeps the vectors 
parallel at each infinitesimal step is called \textit{parallel transport}. 

Given a vector $\mathbf{A}$ at a point $P$, there exists one and only one vector 
$\mathbf{A}^{\prime}$ at point $P^{\prime}$ that can be constructed by parallel transport of $\mathbf{A}$ from $P$ to $P^{\prime}$. 
In Euclidean space, the summation and subtraction of vectors at different points are defined with the help of parallel transport and the triangle rule. The condition of parallel transport may therefore be written as
\begin{equation}
\delta \mathbf{A} = \mathbf{A}^{\prime} - \mathbf{A} = 0,
\label{PT1}
\end{equation}
where $\delta$ stands for an infinitesimal displacement. Equation (\ref{PT1}) is also valid for finite 
displacement.

Parallel transport has a natural description in analytic geometry. The Cartesian coordinate system is 
defined by basis vectors $\mathbf{e}_{i}$ constructed at an arbitrary point $P$ and transported parallel 
to other points:
\begin{equation}
\delta \mathbf{e}_i = \mathbf{e}_i^{\prime} - \mathbf{e}_i = 0.
\label{PT2}
\end{equation}
Contravariant coordinates of a vector $\mathbf{A}$ are fixed by the decomposition 
$\mathbf{A}=A^{i}\mathbf{e}_{i}$; covariant coordinates are scalar products 
$A_{i} = \mathbf{A} \cdot \mathbf{e}_{i}$. Because the basis is orthonormal 
$\mathbf{e}_{i}\cdot \mathbf{e}_{j}=\delta _{ij}$, $\delta_{ij} = \delta^{ij}$, 
$A^{i}=\delta^{ij} A_{j}=A_{i}$ and $\mathbf{A} \cdot \mathbf{B}=A_{i}B^{i}$.

Parallel transport does not change the Cartesian coordinates of vectors: 
\begin{eqnarray}
\delta A_i &=& \delta (\mathbf{e}_i \cdot \mathbf{A}) \nonumber \\
           &=&  \delta \mathbf{e}_i \cdot \mathbf{A} + \mathbf{e}_i \cdot \delta \mathbf{A} = 0.
\label{RIEME33}
\end{eqnarray}
As a consequence, the scalar product of vectors remains invariant, 
\begin{equation}
\delta (\mathbf{A} \cdot \mathbf{B}) = \delta A_i B^i +A_i \delta B^i =0. 
\end{equation}

\subsection{Riemannian space}

Riemannian space is locally Euclidean space. At every point $P$ one may find a coordinate 
system in which the metric takes Euclidean form and at every point $P^{\prime}$ of a neighborhood 
of $P$ the deviation of  the metric from the Euclidean metric is of the second order in the
distance between $P$ and $P^{\prime}$. 

The notion of straight lines makes sense locally. In a locally Cartesian coordinate system at $P$, 
straight lines that pass through $P$ are parameterized by $\mathbf{x} = \mathbf{v}\Delta \ell + O(\Delta \ell^3)$, 
where $\Delta \ell$ is the distance from $P$. \cite{LIFT} 
Geodesic curves look locally like straight lines.
Straight lines determine the shortest lengths between two points, so geodesic curves have the shortest lengths 
locally and globally.

The concept of parallel transport of vectors also makes sense locally. In a local Cartesian coordinate 
system, parallel transport is defined by equation (\ref{PT1}) to first order in displacement. 
Globally, parallel transport refers to transformation along a curve, where a vector undergoes local parallel transport at every step. In Riemannian space, parallel transport preserves the scalar product and as a consequence lengths and angles between vectors. 

During parallel transport along a straight line in Euclidean space, the angle the vector $\mathbf{A}$ makes 
with the path remains fixed. Such a property holds locally in Riemannian space. 
Together with the requirement of fixed length, this uniquely defines 
parallel transport in two-dimensional Riemannian space. As a consequence, the angle remains fixed globally 
along the entire geodesic.

In higher dimensions there still exists a freedom to rotate $\mathbf{A}$ around a vector tangent to the path. Under parallel transport in Euclidean space and locally in Riemannian space, $\mathbf{A}$ remains within the initial two-dimensional plane spanned by $\mathbf{A}$ and the vector $\mathbf{v}$ tangent to the path at $P$. 
This property forbids arbitrary rotations and uniquely defines parallel transport in Riemannian spaces of higher dimensions.
\cite{ROTA}

\subsection{Riemannian space as hypersurface of Euclidean space}

Riemannian space can be imbedded into a Euclidean space $\mathbb{E}$ of higher dimension and treated 
as a hypersurface $\Sigma \subset \mathbb{E}$. 
A global Cartesian coordinate system of hyperplane $\Pi(P)$ tangent to hypersurface $\Sigma$ at point $P$ 
may be chosen as a local Euclidean coordinate system of $\Sigma$ at $P$.
In the neighborhood of $P \in \Sigma$, metric relations and algebraic operations with tangent vectors coincide 
within the hyperplane and hypersurface to first order in distance from $P$. This gives the freedom 
to operate with the simplest geometric objects in Riemannian space locally as though they are 
in Euclidean space.

The tangent components of the vectors of the embedding Euclidean space satisfy Eqs. (\ref{PT1}) and (\ref{PT2}). 

Vectors of Euclidean space $\mathbb{E}$ have components orthogonal to $T_P \Sigma$. The conditions for parallel 
transport in $\Sigma$
do not limit a change of these components. For any extension of the definition, a parallel transport in $\Sigma$ 
of a vector $\mathbf{A} \in T_P \mathbb{E}$ is \textit{not} the parallel transport in $\mathbb{E}$ in general. 
From the standpoint of $\mathbb{E}$, a precession of $\mathbf{A}$ occurs.

Let the vectors $\mathbf{e}_i$ form a local Cartesian basis in tangent space at $P$. 
The same vectors are basis vectors of $\Pi(P)$. In the infinitesimal vicinity of $P \in \Sigma$, parallel transport conditions (3) and (4) 
for vectors $\mathbf{A} \in T_P \mathbb{E}$ may be written as
\begin{eqnarray}
\mathbf{e}_i \cdot \delta \mathbf{A}   &=& 0, \label{PT3} \\
\mathbf{e}_i \cdot \delta \mathbf{e}_k &=& 0. \label{PT4}
\end{eqnarray}
The first order variance of the tangent components of $\mathbf{A}$ with respect to the displacement 
vanishes. Equation (\ref{PT4}) shows that in a local Cartesian coordinate system the Cristoffel symbols of the 
hypersurface vanish also.

The normal component of the variation of $\mathbf{A}$ is not constrained by Eq. (\ref{PT3}). 
Fermi-Walker transport \cite{FERM,WALK,MOLL} provides a condition for the variation of the normal 
component of the vectors as well.
In the problems of Foucault's pendulum and Thomas precession, the normal components 
of vectors vanish identically. In such a situation, Eq. (\ref{PT3}) is kinematically complete.

\begin{figure}[htp]
\centering
\includegraphics[totalheight=0.13\textheight,viewport=0 0 540 250,clip]{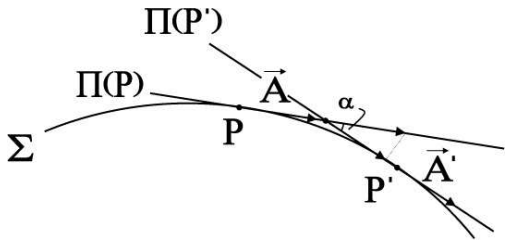}
\caption{Parallel transport of vector $\mathbf{A}$ between infinitesimally close points $P$ and $P^{\prime}$. $\Sigma$ is hypersurface,
$\Pi(P)$ and $\Pi(P^{\prime})$ are hyperplanes tangent to $\Sigma$ 
at $P$ and $P^{\prime}$, respectively. $\mathbf{A}$ is transported first 
within $\Pi(P)$ to an intersection point of $\Pi(P)$ and $\Pi(P^{\prime})$. 
Then $\mathbf{A}$ is projected onto $\Pi(P^{\prime})$ and is transported 
within $\Pi(P^{\prime})$ to $P^{\prime}$. 
The distance between $P$ and $P^{\prime}$
is of the first order in $\alpha$, whereas the variation of length 
of the vector $\delta |\mathbf{A}| = (1 - \cos \alpha)|\mathbf{A}|$
is of the second order in $\alpha$. In the continuum limit, 
$|\mathbf{A}|$ remains fixed.
}
\label{figp}
\end{figure}

Parallel transport within a hypersurface from $P$ to $P^{\prime}$, when $P^{\prime}$ is infinitesimally close to $P$, 
can therefore be considered as being equivalent to parallel transport of a vector within $\Pi(P)$ and its 
subsequent projection onto $\Pi(P^{\prime})$, as shown on Fig. \ref{figp}. 
The result is independent of the point of the intersection of $\Pi(P)$ and
$\Pi(P^{\prime})$ at which the projection is made. Once parallel transport is defined under infinitesimal 
displacements, it is defined in the integral sense.

\vspace{2mm}

Equations (\ref{PT3}) and (\ref{PT4}) lead to the conclusion that, if the basis vectors and tangent vector $\mathbf{A}$ 
simultaneously undergo parallel transport along a geodesic curve, the coordinates of $\mathbf{A}$ measured in the 
local basis remain fixed. This statement follows formally from Eqs. (\ref{RIEME33}) - (\ref{PT4}) 
and the expansion $\mathbf{A} = A^i \mathbf{e}_i$:
\begin{eqnarray}
\delta (\mathbf{e}_i \cdot \mathbf{A}) &=& \delta \mathbf{e}_i \cdot \mathbf{A} + \mathbf{e}_i \cdot \delta \mathbf{A} \nonumber \\
                                       &=& A^k \mathbf{e}_k \cdot \delta \mathbf{e}_i + \mathbf{e}_i \cdot \delta \mathbf{A} = 0.
\end{eqnarray}

In the spherical basis of a sphere $\mathbb{S}^2$, the basis vectors $\mathbf{e}_{\vartheta}$ and $\mathbf{e}_{\varphi}$ 
at different $\vartheta$ and fixed $\varphi$ are related by parallel transport. As a result, the 
parallel transport of $\mathbf{A}$ along 
the meridians of the surface does not change the local coordinates of $\mathbf{A}$. 
In particular, the orientation of the plane of the pendulum relative to the remote stars can naturally 
be defined in the local coordinate system of an observer at the North Pole 
by invoking the parallel transport of the pendulum along the meridians. Such transport does not change $\varphi_E$.

The synchronization of coordinate systems in the special theory of relativity assumes that the basis 
vectors of systems $S^{\prime}$ are obtained by boost transformations of the basis vectors of
some preferred coordinate system $S$. 
The statements that the one-parametric family of boosts in a two-dimensional plane of $T_x \mathbb{M}$
defines some geodesic in the hyperboloid $u^2=1$ of relativistic velocities and that the 
boost of basis vectors is a parallel transport along a geodesic are proved in Appendix. 
As a consequence, we note that parallel transport of a polarization
four-vector $a$ along a geodesic does not change the local coordinates of $a$.

The general mathematical formalism required for the description of Riemannian spaces can be 
found in the classical textbooks \cite{arnold,LANDLI,MOLL}. 

A sphere embedded in a three-dimensional Euclidean space, $\mathbb{R}^3$, and a hyperboloid of physical relativistic 
velocities embedded in a four-dimensional space of relativistic velocities, 
$T_x \mathbb{M}$, are simple enough, 
allowing for a description of parallel transport using elementary tools.

\section{Rotation of the swing plane of Foucault's pendulum}

\subsection{Dynamic conditions}

Pedagogical introductions to the dynamics of Foucault's pendulum are widely available. 
We focus on those aspects of the dynamics which are 
closely related to the geometry of the problem (see also \cite{SOME,HART,WILC,JENS,BERR}).

The suspension point of the pendulum traces out a circular path. The pendulum experiences the Coriolis 
force. The reaction force of the pendulum wire resists gravity \cite{COM4}. 

Gravity points towards the center of the Earth while the force of the wire depends on the motion 
of the pendulum. The wire compensates for the free-fall acceleration $\mathbf{g}$ and 
creates a restoring force when the bob deviates from equilibrium.

\begin{figure}[htp]
\centering
\includegraphics[totalheight=0.35\textheight,viewport=135 149 474
591,clip]{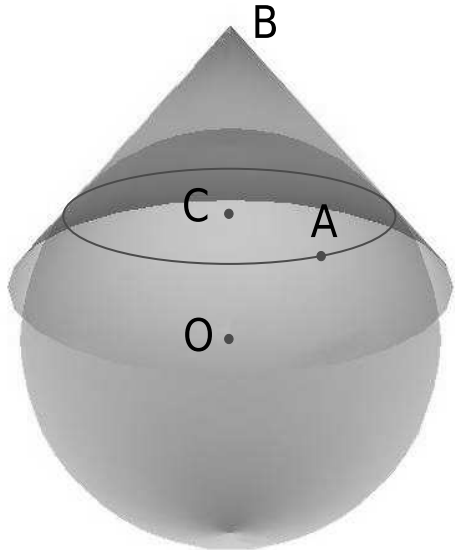}
\caption{The Earth is shown as an ideal sphere centered at $O$, with the radius 
$|\vec{OA}|$. The Foucault's pendulum is placed initially at the point $A$
on the circle of fixed polar angle $\vartheta = \angle BOA$. During one sidereal
day, the Foucault's pendulum completes one period of rotation. Parallel
transport of the pendulum's plane can be modelled by
considering vector normal to the pendulum's plane as belonging 
to tangent space of the sphere and therefore of the surface of a
cone coaxial with the Earth and tangent to it at the latitude of the pendulum
(the cone looks like a Vietnamese hat). The apex of the cone is at the point $B$. 
The point $C$ is the center of the circle along which the pendulum moves. 
$B$ and $C$ belong to the rotation axis of the Earth.}
\label{fig1}
\end{figure}

The Coriolis force appears in the equations of motion because of the rotation 
of the pendulum in the Earth-bound coordinate system: 
\begin{equation}
\mathbf{F}_C = 2m \mathbf{v} \times  \mbox{\boldmath{$\Omega$}},
\label{coriolis}
\end{equation}
where $m$ and $\mathbf{v}$ are the mass and velocity of the bob, and
$\mbox{\boldmath{$\Omega$}}$ is the rotation frequency of the Earth. 

At the Equator, $\mbox{\boldmath{$\Omega$}}$ and $\mathbf{v}$ belong to tangent space, 
so that $\mathbf{F}_C$ is collinear with $\mathbf{g}$. In such a case, gravity, the wire, 
and the Coriolis force do not change the pendulum's state. In the coordinate system of the Earth, 
the swing plane of the pendulum does not rotate 
with respect to the direction of its motion as the pendulum moves along the Equator. 
This property holds for motion along any arc of a great circle of the Earth.

Now, we are in a position to reformulate the \textit{dynamical evolution} of the pendulum as a purely 
geometrical problem of \textit{parallel transport} of the pendulum swing plane.

For the geodesic approximation of the path of the pendulum, one may choose a set of small arcs 
of great circles of the Earth. Along such arcs the state of the pendulum is conserved, 
as its swing plane does not rotate relative to the path direction. The swing
plane experiences \textit{motion by inertia}. However, relative to
the circle of a polar angle $\vartheta$, the swing plane rotates because
geometry of the Earth's surface is non-Euclidean. In the continuum limit when the
lengths of the arcs tend to vanish, one can reconstruct the initial circle of the 
polar angle $\vartheta$ and calculate the rotation angle 
of the plane of the pendulum swings during one sidereal day.

The Earth has a nonvanishing flattening $f \approx 1/300$ ($f=(a-c)/a$ 
where $a$ is the equatorial radius and $b$ is the polar radius of an ellipsoid). 
We treat Earth as an ideal sphere and neglect small deviations from the 
parallel transport connected with that
ellipsoidal form.

\subsection{Tangent cone method in Foucault's pendulum problem}

In Fig. \ref{fig1} the Earth is pictured covered from the Northern Hemisphere by a circular cone with apex $B$. 
The cone is tangent to the Earth at a polar angle $\vartheta = \angle BOA$. A vector orthogonal to the swing plane 
moves together with the pendulum, as shown on Fig. \ref{fig1}. Such a vector belongs to the tangent spaces of 
the sphere and the cone. We set the Earth's radius equal to unity and obtain 
\begin{eqnarray}
\vec{OA} &=& ( \mathbf{n} \sin \vartheta, \cos \vartheta),  \label{11} \\
\vec{CA} &=& ( \mathbf{n} \sin \vartheta, 0),  \label{12} \\
\vec{BA} &=& ( \mathbf{n} \sin \vartheta, - \frac{\sin \vartheta ^2}{\cos
\vartheta}),  \label{13}
\end{eqnarray}
where $\mathbf{n} = (\cos \varphi, \sin \varphi)$ is a unit vector in the equatorial plane. In order to
find $\vec{BA}$, we write $\vec{BA} = ( \mathbf{n} \sin \theta, z)$ and fix $%
z$ using the orthogonality 
\begin{equation}
\vec{BA} \cdot \vec{OA} =0.
\label{14}
\end{equation}

The metric induced by the Euclidean space $\mathbb{R}^3$ on the cone is Euclidean. Indeed,
one may choose a coordinate system on the cone 
$(\rho,\varphi)$ where $\rho$ is the Euclidean distance from $B$ and $\varphi$ is the azimuth angle
defined earlier. The infinitesimal distance between two points on the cone
is equal to 
\begin{equation}
d\ell^2 = d\rho^2 + \cos^2 \vartheta \rho^2 d\varphi^2.  \label{EUCL}
\end{equation}
By a change of the variable $\varphi \rightarrow \varphi/\cos\vartheta$
(recalling that $\vartheta$ is constant), the metric of the cone becomes that
of a plane in the polar coordinate system.

\begin{figure}[htp]
\centering
\includegraphics[totalheight=0.25\textheight,viewport=119 217 470
581,clip]{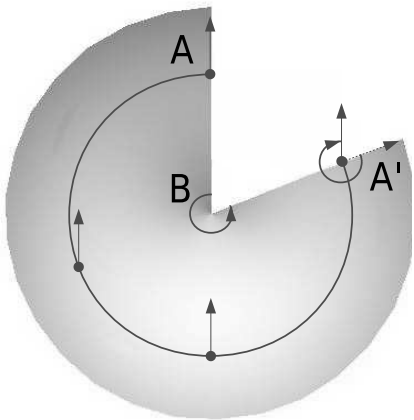}
\caption{ The cone of Fig. 1 is cut along the line $BA$, unfolded preserving
the distances and placed on the plane to make parallel transport of the
tangent vectors visual. Foucault's pendulum is initially at the point $A$.
It rotates with the Earth counterclockwise and completes one period 
when arrives at $A$. On the plane, however, it arrives to the point $A^{\prime} \neq A$ which
is physically the same as $A$. The path on the plane is not closed and the
rotation angle $\varphi_{E}$ shown by the oriented arc around the point $%
A^{\prime}$ is smaller than $2\pi$ in absolute value. $\varphi_{E}$
determines the rotation angle of the swing plane relative to the Earth. 
The points $A$ and $B$ are the same as in Fig. 1. }
\label{fig2}
\end{figure}

It is thus possible to cut the cone along the straight line $BA$, unfold
it and place it on a plane (Fig. \ref{fig2}). The metric on the cone remains
Euclidean. The distances between points on the cone and angles do not change.

Parallel transport in Euclidean space is, however, simple and evident. This is shown in
Fig. \ref{fig2} for a vector tangent at $A$ to the meridian. 

The pendulum rotates counterclockwise with the Earth, whereas the plane of the
swings rotates clockwise in the
direction of the rotation of the stars. 
In the Earth's coordinate system, the
rotation angle $\varphi_{E}$ is negative. Its value is determined by the
ratio between the length of the arc $AA^{\prime}$ along the path of the pendulum, 
$2\pi |\vec{CA}|$, and the radius $|\vec{BA}|$ of the circle shown 
in Fig. \ref{fig2}. 
The length of the arc $AA^{\prime}$ is the length of a circle centered at $C$ in Fig. \ref{fig1},
$|\vec{CA}|$ is its radius and $|\vec{BA}|$ is the slant height of the cone shown in Fig. \ref{fig1}.
The lengths of the vectors defined by Eqs. (\ref{12}) and (\ref{13}) are 
$|\vec{CA}| = \sin \vartheta$ and $|\vec{BA}| = \tan \vartheta$,
and so we obtain
\begin{equation}
\varphi_{E} = -\frac{2\pi |\vec{CA}|}{|\vec{BA}|} = - 2\pi \cos \vartheta.
\end{equation}
Relative to the remote stars, the rotation angle is that of Eq. (\ref
{ODINFOUC}). 

Equation (\ref{ODINFOUC}) is derived for the Northern Hemisphere where $%
\vartheta \le \pi/2$. Its validity extends to the Southern Hemisphere. To
prove it, one has to flip Fig. \ref{fig1} horizontally and repeat the
arguments.

\section{Thomas precession}

Figure \ref{fig3} shows the three-dimensional projection of the space of relativistic velocities 
and the hyperboloid of physical relativistic velocities 
\begin{equation}
u \cdot u = 1.
\end{equation}
The scalar product is calculated using the Minkowski metric $g_{\mu \nu} = 
\mathrm{diag}(1,-1,-1,-1)$. The particle is placed initially at the point $A$
of the hyperboloid. It has a four-velocity $u = (\gamma, \gamma \mathbf{v}%
/c)$, where $\mathbf{v}$ is the three-dimensional velocity.

\subsection{Tangent space of relativistic velocities}

\subsubsection{Polarization vector}

The polarization vector of a particle with spin $s$ is defined in its rest frame
as the expectation value of the spin operator $\hat{\mathbf{s}}$: 
\begin{equation}
\mathbf{a} = \frac{1}{s}<\hat{\mathbf{s}}>.
\end{equation}
Transformation properties of three-dimensional vectors under Lorentz transformations 
are uncertain. The polarization of a relativistic spin-$s$
particle must be described by a four-dimensional vector $a$. In its rest
frame, polarization can be defined as follows 
\begin{equation}
a=(0,\mathbf{a}),  \label{403}
\end{equation}
with $\mathbf{a}^2 = -a^2 = 1$ in pure states and $\mathbf{a}^2 = -a^2 < 1$
in mixed states.

In the rest frame of the particle, 
\begin{equation}
u=(1,\mathbf{0}),  \label{404}
\end{equation}
and therefore 
\begin{equation}
u \cdot a=0.  \label{405}
\end{equation}
The scalar product is invariant under Lorentz transformations, so Eq. (\ref{405})
holds in all inertial coordinate systems.

The equivalence principle implies that all phenomena in the local inertial
frame are the same as in the global inertial frame. The representation (\ref
{403}) therefore holds in the inertial co-moving coordinate systems of
particles which move with some non-zero acceleration. The admissible variations of $a$ in
the local inertial frame are restricted to rotations of $\mathbf{a}$.

It is possible to arrive at Eq. (\ref{405}) differently. Using the angular momentum tensor $M_{jk}$ and the four-momentum $p_l = mu_l$, 
one can construct a Pauli-Lubanski vector 
\begin{equation}
J^i = \frac{1}{m}\varepsilon^{ijkl} M_{jk}p_l,  \label{PL}
\end{equation}
which is proportional to the polarization four-vector $J^i = sa^i$. In order to show
that, one may choose the rest frame $p=(m,\mathbf{0})$ where the orbital
momentum is zero, while the spin part of $M_{jk}$ contributes to $J^i$.
Equation (\ref{405}) holds since $\varepsilon^{ijkl}$ is totally
antisymmetric ($\varepsilon_{0123} = +1$).

From the geometrical point of view, Eq. (\ref{405}) implies that $a$ belongs
to the tangent space of $T_x \mathbb{M}|_{u^2 = 1}$, i.e., to tangent space of
the hyperboloid $u^2 = 1$ of relativistic velocity space $T_x \mathbb{M}$.

There are two contexts of parallel transport of the precession to consider. One is parallel 
transport along the spacetime path of a spinning particle. This spacetime path is a helix in 
$\mathbb{M}$. The other is parallel transport along a latitudinal circle 
of the hyperboloid. 

The polarization vector $a$ implicitly depends on the four-dimensional label $x \in \mathbb{M}$ and the three-dimensional 
label $u \in T_x \mathbb{M}|_{u^2 = 1}$.

The two polarization vectors $a$ and $a^{\prime}$ moving with four-velocities $u \in T_x \mathbb{M}|_{u^2 = 1}$ 
and $u^{\prime} \in T_{x^{\prime}} \mathbb{M}|_{u^2 = 1}$ at points $x \in \mathbb{M}$ 
and $x^{\prime} \in \mathbb{M}$ can be compared first when the four-velocities are equal: $u = u^{\prime}$.
Two observers at $x \ne x^{\prime}$ moving with velocities $u = u^{\prime}$ belong to the same 
inertial coordinate system, up to some rotation. The second observer turns the axes of his frame 
in the direction of the axes of the first observer's frame. They communicate and inform each other of the coordinates of $a$ and 
$a^{\prime}$ that they measure. Such a comparison is formally equivalent to parallel 
transporting the vectors $a$ and $a^{\prime}$ in Minkowski spacetime
as if these vectors belonged to $T_x \mathbb{M}$, although according to Eq. (\ref{405}), they belong to the tangent space of $T_x \mathbb{M}|_{u^2 = 1}$.

For $x=x^{\prime}$ and $u \neq u^{\prime}$, the polarization vectors are compared by using the scheme of parallel transport in the
physical relativistic velocity space along the geodesic connecting $u$ and $u^{\prime}$.

For $x \neq x^{\prime}$ and $u \neq u^{\prime}$, the vector $a^{\prime}$ is parallel transported $(x^{\prime},u^{\prime}) \to (x,u^{\prime})$ in $\mathbb{M}$, 
then it is parallel transported $(x,u^{\prime}) \to (x,u)$ along the geodesic in $T_x \mathbb{M}|_{u^2 = 1}$ and after that $a^{\prime}$ is compared to $a$. 

Since parallel transport in $\mathbb{M}$ does not change the vector coordinates, and, hence, parallel transports
in $\mathbb{M}$ and $T_x \mathbb{M}|_{u^2 = 1}$ commute, the result is independent of the order of the operations.
Thus, the polarization vectors can be regarded as vectors
$T_x \mathbb{M}|_{u^2 = 1}$ for any chosen value of $x$, e.g., $x = (0,\mathbf{0})$ and can
be characterized by a four-velocity $u \in T_x \mathbb{M}|_{u^2 = 1}$ only.
As a result, an observer based in some reference frame, e.g., at the origin, is able to analyze and make consistent
conclusions on the character of the spin precession of a particle moving with any velocity and acceleration.
So, while a spinning particle follows a helix in Minkowski spacetime, just one observer
draws conclusions about the precession. 

In quantum mechanics, the uncertainty relations do not allow a simultaneous measurement of
the coordinates and velocity. This restriction does not pose constraints on the formalism, since the 
localization of particles is not important, only shifts in the velocity space contribute to the precession.

\subsubsection{Angular momentum of gyroscope}

The angular momentum of gyroscope, $\mathbf{L}$, is a three-dimensional vector
in its rest frame. The same arguments as for the polarization vector lead us to
introduce a four-vector to characterize it in a relativistically invariant
fashion. The easiest way to do that is to start from the Pauli-Lubanski vector (%
\ref{PL}). We arrive at the same conclusions as for the polarization vector,
namely, that the representation $J=(0,\mathbf{L})$ takes place in the rest
frame and $J$ is tangent to the space of physical relativistic velocities, i.e., 
$u \cdot J = 0$.
In view of the above similarity, a gyroscope is often treated as a
mechanical analog of a spinning electron.

Under the action of an external force applied to the center of mass in a chosen direction, 
the gyroscope is parallel transported from one inertial coordinate system to another. For
this reason, gyroscope is convenient to represent coordinate axes of inertial coordinate 
systems.

\subsubsection{Four-acceleration of particle}

By taking the derivative of the equation $u^2 = 1$, one gets $w \cdot u=0$. The
acceleration $w = du/ds$ therefore belongs to the tangent space of $T_x\mathbb{M}|_{u^2 = 1}$ also.

\subsection{Tangent cone method in spin precession problem}

Any sphere and any hyperboloid can be covered by coaxial circular cones. 
We construct a cone tangent to the hyperboloid at point $A$ and along a circle of fixed $\gamma$ as well, 
i.e., along the particle trajectory in the space of physical 
relativistic velocities. On the circular orbit, the tangent spaces of the hyperboloid and the cone coincide. 
The polarization vector $a$ belongs to those tangent spaces. 

\begin{figure}[htp]
\centering
\includegraphics[totalheight=0.30\textheight,viewport=126 210 489
583,clip]{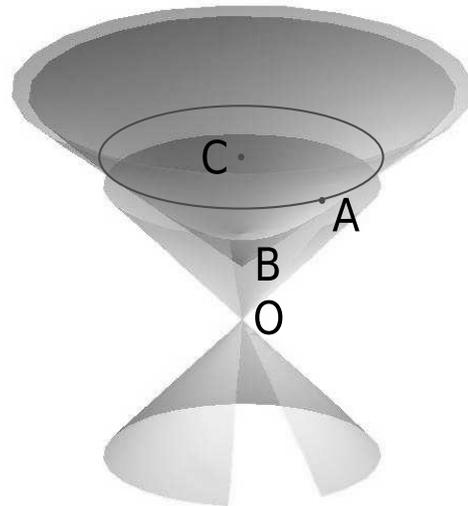}
\caption{A three-dimensional projection of the relativistic velocity space.
The hyperboloid is a set of physical relativistic velocities with 
$u_{0} = + \sqrt{1 + \mathbf{u}^2}$, $\mathbf{u} = \mathbf{n}\gamma v/c$ are three-dimensional vectors. 
The four-vector $\vec{OA} = (\gamma, \mathbf{n}\gamma v/c)$ shows the initial four-velocity of the particle on a circular
orbit of constant $\gamma$. $\vec{BA}$ belongs to tangent space of the
hyperboloid, since it is orthogonal to $\vec{OA}$. 
The cone with the apex $B$ is tangent to the hyperboloid at particle trajectory (it looks like a flipped Vietnamese hat). The light cone with the apex $O$ is shown also. }
\label{fig3}
\end{figure}

Let $B$ be the apex of the cone. The vectors of Fig. \ref{fig3} have the form 
\begin{eqnarray}
\vec{OA} &=& (\gamma, \mathbf{n} \gamma v/c), \label{21} \\
\vec{CA} &=& (0 , \mathbf{n} \gamma v/c), \label{22} \\
\vec{BA} &=& (\gamma v^2/c^2, \mathbf{n} \gamma v/c), \label{23}
\end{eqnarray}
where $\mathbf{n}$ is a unit vector in the rotation plane of the particle.
In order to fix $\vec{BA}$, one can set $\vec{BA} = (y, \mathbf{n} \gamma
v/c)$ and determine $y$ from the orthogonality 
\begin{equation}
\vec{BA} \cdot \vec{OA} =0.
\label{24}
\end{equation}
Equations (\ref{21}) - (\ref{24}) are similar to equations (\ref{11}) - (\ref{14}). 

The vector $\vec{BA}$ is tangent to the hyperboloid. In the course of the parallel
transport, its rotation angle coincides with the rotation angle of the
polarization vector.

The induced metric on the cone can be obtained as follows: the points 
of the cone are characterized by vectors
\begin{equation}
\vec{BX} = \rho (v/c, \mathbf{n}),  \label{VECT}
\end{equation}
where $\mathbf{n}=(\cos \phi, \sin \phi,0)$ lies in the rotation plane 
$(x,y)$. The vectors $\vec{BX}$ are obtained by dilatation 
and rotation of $\vec{BA}$ around the $z$ axis. We vary $\rho$ and $\phi$ 
at fixed $\gamma$. The variation of $\vec{BX}$ is tangent to the cone and 
can be represented by 
\begin{eqnarray}
d\vec{BX} &=& \frac{\partial \vec{BX}}{\partial \rho}d\rho + \frac{\partial \vec{BX}}{\partial \phi}d\phi \nonumber \\
          &=& d\rho (v/c, \mathbf{n}) + \rho (0, d \mathbf{n}), \label{DV}
\end{eqnarray}
where $d\mathbf{n} = (- \sin \phi, \cos \phi,0)d\phi$, so that $\mathbf{n} \cdot d\mathbf{n} = 0$. 

The length of the four-vectors is given by 
\begin{equation}
|\vec{XY}| = \sqrt{- \vec{XY} \cdot \vec{XY}}.
\label{dist}
\end{equation}
Space-like four-vectors have real lengths. 

The infinitesimal length between two points on the cone is determined from the 
scalar product $d\ell^{2} = - d\vec{BX} \cdot d\vec{BX}$. 
Using Eq. (\ref{DV}), we obtain
\begin{equation}
d\ell^{2} = d\rho^{2}/\gamma^{2} + \rho^{2} d\phi^{2}.  \label{METR}
\end{equation}
In order to bring $d\ell^{2}$ to a form identical to that of Euclidean space, we scale 
$\phi \to \phi/\gamma$ and $\rho \to \gamma \rho$ and arrive at 
\begin{equation}
d\ell^{2} = d\rho^{2} + \rho^{2} d\phi^{2}.  
\label{METR-2}
\end{equation}

\begin{figure}[htp]
\centering
\includegraphics[totalheight=0.24\textheight,viewport=131 229 467
571,clip]{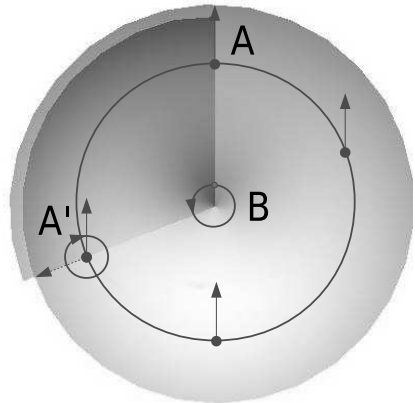}
\caption{
The cone of Fig. \ref{fig3} with the apex at $B$ was cut along $BA$, unfolded with preserving the intervals and placed on the plane.
During one turn of the particle from $A$ to $A^{\prime}$ along the circular orbit its polarization vector rotates by $|\phi_{E}| > 2\pi$ 
in the co-moving coordinate system. The points $A$ ($=A^{\prime}$) and $B$ are the same as in Fig. \ref{fig3}. }
\label{fig4}
\end{figure}

The metric induced on the cone by Minkowski spacetime is thus Euclidean. 
Equation (\ref{METR-2}) gives it in a polar coordinate system.
The cone of Fig. \ref{fig3} can therefore be cut along $BA$, unfolded to preserve
the distances and the angles and placed on a plane as shown on Fig. \ref{fig4}.

The particle rotates counterclockwise, while its polarization vector rotates
clockwise. In the co-moving coordinate system, the rotation angle $\phi_{E}$ 
is negative. Its value is determined by the ratio between the length of the
arc $AA^{\prime}$ along the path of the particle trajectory i.e. $2\pi |\vec{CA}|$ and the
circle of radius $|\vec{BA}|$.

The path represents a circle of fixed latitude in the hyperboloid of physical 
relativistic velocities as shown on Fig. \ref{fig3}. Its length 
is fixed by the radius $|\vec{CA}|$. 
On the plane of Fig. \ref{fig4}, the angle covered by the arc $AA^{\prime}$ 
along the path is greater than $2\pi$.

The lengths between the points $A$, $B$ and $A$, $C$ can be found with the 
help of Eqs. (\ref{22}) and (\ref{23}):
\begin{eqnarray}
|\vec{CA}| &=& \gamma v/c, \\
|\vec{BA}| &=& v/c.
\end{eqnarray}
$\vec{CA}$ and $\vec{BA}$ are space-like four-vectors and their lengths make
physical sense. Finally, we obtain 
\begin{equation}
\phi_{E} = - \frac{2\pi |\vec{CA}|}{|\vec{BA}|} = - 2\pi \gamma.
\end{equation}
In an inertial coordinate system the origin of which $x=y=z=0$ coincides with the 
axis of a spiral trajectory of the particle in $\mathbb{M}$,
the rotation angle $\phi_{S} = 2\pi + \phi_{E}$ takes the form of Eq. (\ref
{ODINTHOM}). 

In the space $T_x \mathbb{M}$, such a coordinate system is in the vertex $u = (1,\mathbf{0})$
of the hyperboloid  $u^2 = 1$. From the standpoint of the problem of Foucault's pendulum, 
it can be viewed as an analogue of a local coordinate system of an observer on the North Pole.

Meridians on a sphere are formed by a trace of the point $(0,0,1)$ under rotation in the planes of vectors $(0,0,1)$ and 
$(\cos \varphi, \sin \varphi,0)$, where $\varphi$ is the azimuthal angle numbering the meridians. Analogous 
to meridians are orbits in $T_x \mathbb{M}$ formed by a trace of the point $u = (1,\mathbf{0})$ under boost
transformations in the planes of four-vectors $(1,\mathbf{0})$ and $(0,\mathbf{n})$, where $\mathbf{n}$ is the
unit vector numbering the orbits. Orbits of the point $(1,\mathbf{0})$ are geodesics as well as meridians on a sphere 
(see Appendix).

Equation (\ref{ODINFOUC}) can be converted to Eq. (\ref{ODINTHOM}%
) by replacing $\vartheta \to i\theta$ which corresponds formally to the replacement
of rotations by Lorentz transformations. 

In the co-moving coordinate system, the particle experiences a centrifugal force, the Coriolis
force, and the reaction force. These forces act to keep the particle at rest.
Their cumulative effect on the spin rotation, however, cannot be estimated
without specifying the Lorentz nature of the reaction force.

The circular orbit of the particle can be approximated by a set of small segments of geodesics 
of the hyperboloid. Such segments correspond to boost transformations which induce parallel 
transport of the polarization vectors as discussed in Section II.
Geodesics of the hyperboloid are analogues of the arcs of great circles
of a sphere. In the continuum limit, geodesic approximations of the path become a circle 
of constant $\gamma$. Parallel transport along the path gives a geometric, universal component of the spin precession. 

In the external Lorentz-scalar potentials, the reaction force does not 
contribute to the spin precession, so the Thomas precession effect gives the complete answer. 
This can be interpreted to mean that Lorentz-scalar potentials do not generate torques. 

For external Lorentz-vector potentials, the spin of the particles does 
experience torques which result in the so-called Larmor precession \cite{LALI3}. 
The latter case is that of ordinary \cite{THOM26,LALI3} and $\mu$-meson and hyperon 
exotic atoms \cite{GIAN,BATT,omega}. 
The Thomas and Larmor effects thoroughly determine the spin precession in agreement 
with the Bargmann-Michel-Telegdi equation \cite{BMT,omega}. 

Due to purely kinematic nature, Thomas spin precession affects spectroscopy and static 
characteristics of nuclei \cite{LALI3} and hadrons \cite{MIKR,KOBZ}.

\section{Conclusion}

A simple method employing tangent cones that has been used earlier to illustrate the effect of curvature on the 
rotation of vectors transported along the surface of a sphere \cite{arnold} and 
to describe the rotation angle of the swing plane of Foucault's pendulum 
\cite{SOME,HART} has been extended to describe the effect of Thomas precession of 
relativistic particle moving along a circular orbit. 

In the problems of Foucault's pendulum and Thomas spin precession, vectors characterizing the system are not
influenced by an external rotating moment and evolve by inertia, undergoing parallel transport.
A close analogy between parallel transport of vectors along the surface of a sphere in three-dimensional Euclidean space 
and the hyperboloid $u^2 = 1$ in the space of relativistic velocities has been exploited.
In both cases, the evolution is reduced to parallel transport in the usual Euclidean space represented by the tangent cone surface.

The tangent cone method facilitates a straightforward derivation of Eq. (\ref{ODINTHOM}) and 
interpretation of Thomas spin precession using our intuition of parallel transport in Euclidean space.

\appendix

\section{Geodesics in relativistic velocity space}

Here, we prove two statements made at the end of Section II.

\textit{1}. The points $u, u^{\prime} \in T_x\mathbb{M}|_{u^2=1}$ describe two inertial reference frames $S$ and $S^{\prime}$. 
In the frame $S$, $u =(1, \mathbf{0})$. We write $u^{\prime}$ as
\begin{equation}
u^{\prime} = (\cosh \theta, \mathbf{n}\sinh \theta),
\label{HS}
\end{equation}
where $\mathbf{v} = c \mathbf{n} \tanh \theta$ is the velocity of $S^{\prime}$ in the frame $S$, and
$\mathbf{n}$ is the unit vector. Reference frames $S$ and $S^{\prime}$ are related by a boost in the plane $(u,u^{\prime})$.

The metric induced in $T_x \mathbb{M}|_{u^2 = 1}$ is determined by the interval $ds^2 = du \cdot du $ 
and in variables (\ref{HS}) is given by
\begin{equation}
ds^2 = - d\theta^2 - \sinh ^2 \theta d \mathbf{n}^2.
\label{metricM}
\end{equation}

The interval between two points on $T_x \mathbb{M}|_{u^2 = 1}$ is negative,
furthermore, $- ds^2 \geq d\theta ^{2}$. Thus, we find
\begin{equation}
\int^{u^{\prime}}_{u} \sqrt{-ds^2} \geq \theta. 
\label{nice}
\end{equation}
Any deviations from curve (\ref{HS}) that connects points $u$ and $u^{\prime}$ by $\theta$ for fixed $\mathbf{n}$ increase the distance between $u$ and $u^{\prime}$.
Hence, the set of 4-velocities (\ref{HS}) with constant $\mathbf{n}$ determines a geodesic on $T_x \mathbb{M}|_{u^2 = 1}$. On the other hand, this
geodesic is an orbit formed by a trace of the point $u =(1, \mathbf{0})$ under boost transformations in the plane of vectors $(1,\mathbf{0})$ and
$(0,\mathbf{n})$.

\textit{2}. Basis vectors of the tangent space $T_x \mathbb{M}|_{u^2 = 1}$ can be chosen as
\begin{eqnarray}
e_{\theta}     &=& \frac{\partial u^{\prime}}{\partial \theta}    = (\sinh \theta, \mathbf{e}_{r} \cosh \theta), \label{metricM1} \\
e_{\vartheta}  &=& \frac{1}{\sinh \theta}\frac{\partial u^{\prime}}{\partial \vartheta} = (0, \mathbf{e}_{\vartheta}), \label{metricM2} \\
e_{\varphi}    &=& \frac{1}{\sinh \theta \sin \vartheta } \frac{\partial u^{\prime}}{\partial \varphi} = (0, \mathbf{e}_{\varphi}),
\label{metricM3}
\end{eqnarray}
where $\mathbf{e}_{r} = \mathbf{n} \equiv (\sin \vartheta \cos \varphi, \sin \vartheta \sin \varphi, \cos \vartheta)$
and 
\begin{eqnarray}
\mathbf{e}_{\vartheta } &=& \frac{\partial \mathbf{n}}{\partial \vartheta}, \label{basisE2} \\
\mathbf{e}_{\varphi} &=& \frac{1}{\sin \vartheta} \frac{\partial \mathbf{n}}{\partial \varphi}
\label{basisE3}
\end{eqnarray}
are the basis vectors of the spherical reference frame in $\mathbb{R}^3$. We note that $e_{\alpha} \cdot e_{\beta} = - \delta_{\alpha \beta}$ along the geodesic.

For a displacement $\theta \to \theta + \delta \theta$ corresponding to a boost
in the direction $\mathbf{n}$, basis vectors (\ref{metricM1}) - (\ref{metricM3}) change. Their
variations satisfy the conditions
\begin{equation}
e_{\alpha} \cdot \delta e_{\beta} = 0,
\label{ns}
\end{equation}
which are the parallel transport conditions according to (\ref{PT4}).


\begin{thebibliography}{99}
\bibitem{LIFT} {If a lift is falling down with the free-fall acceleration,
trajectories of probing particles that pass through the center-of-mass of the lift (at point $P$) deviate from 
straight lines to order $O(\Delta \ell^3)$, because away from $P$ they feel gravitational force.}
\bibitem{ROTA} Also, one can consider geodesic curves emanating from $P$ in the plane $\Pi_2$ spanned by $\mathbf{A}$ and the vector $\mathbf{v}$ tangent to the path. Such geodesic curves form a two-dimensional surface $\Sigma_2$ tangent to $\Pi_2$. Under parallel transport from $P$ to $P^{\prime} \in \Sigma_2$, the vector $\mathbf{A}$ remains tangent to $\Sigma_2$. Such a prescription
also forbids rotations and gives the same $\mathbf{A}^{\prime}$ at $P^{\prime}$ with accuracy $O(\Delta \ell^{2})$. 
In the limit of $\Delta \ell \to 0$, parallel transport is well defined.
\bibitem{COM4} {
The centrifugal force brings vector orthogonal to the swings plane of Foucault's pendulum out of 
the tangent space of the surface of the Earth. The effect originates from a slight perturbation 
of the free-fall acceleration $\mathbf{g}$ and can be neglected with accuracy of ${\mbox{\boldmath{$\Omega$}}}^{2}R \sin \vartheta /|\mathbf{g}| \sim 0.003$, where 
$|\mbox{\boldmath{$\Omega$}}| = 2\pi/23.93$ h$^{-1}$, $R = 6371$ km is the Earth's radius, $\sin \vartheta \sim 1$, and $|\mathbf{g}| = 9.81$ m/s$^{2}$. 
}
\end{thebibliography}
\end{document}